\documentclass[%
reprint,
superscriptaddress,
showpacs,
amsmath,amssymb,
aps,
prl,
longbibliography,
]{revtex4-1}

\usepackage{psfrag,graphicx,epsfig,color}% Include figure files
\usepackage{dcolumn}% Align table columns on decimal point
\usepackage{bm}% bold math
\usepackage{natbib}
\usepackage[usenames,dvipsnames,svgnames,table]{xcolor}
\usepackage{subfigure}
\usepackage{rotating}
\usepackage{float}
\usepackage[nice]{nicefrac}
\usepackage{tabularx}
\def\re    {{R_\lambda}}

\def\xx {{\mathbf{x}}}
\def\rr {{\mathbf{r}}}
\def\kk {{\mathbf{k}}}

\def\ww {{\boldsymbol{\omega}}}
\def\SS {{\mathbf{S}}}

\begin{document}

\title{
Non-local amplification of intense vorticity in turbulent flows 
}
%Non-locality and and broken scale similarity of gradient amplification in turbulence

%%% Authors %%%
\author{Dhawal Buaria }
\email[]{dhawal.buaria@nyu.edu}
%\thanks{}
\affiliation{Tandon School of Engineering, New York University, New York, NY 11201, USA}
\affiliation{Max Planck Institute for Dynamics and Self-Organization,
G\"ottingen 37077, Germany} 

\author{Alain Pumir}
\affiliation{Laboratoire de Physique, Ecole Normale Sup\'erieure de Lyon, Universit\'e de 
Lyon 1 and Centre National de la Recherche Scientifique, 69007 Lyon, France}
\affiliation{Max Planck Institute for Dynamics and Self-Organization,
G\"ottingen 37077, Germany} 

\date{\today}% It is always \today, today,

%\thispagestyle{empty}

\begin{abstract}

The nonlinear and nonlocal coupling of vorticity and strain-rate
constitutes a major hindrance in understanding the self-amplification
of velocity gradients in turbulent fluid flows.
Utilizing highly-resolved direct numerical simulations of
isotropic turbulence in periodic domains
of up to $12288^3$ grid points,
and Taylor-scale Reynolds number $\re$ in the
range $140-1300$,
we investigate this nonlocality by decomposing
the strain-rate tensor into local and non-local
contributions obtained through Biot-Savart integration of vorticity
in a sphere of radius $R$.
We find that vorticity is predominantly amplified
by the non-local strain coming beyond a characteristic scale size,
which varies as a simple power-law of vorticity
magnitude. The underlying dynamics %are linear 
preferentially align vorticity with the most
extensive eigenvector of non-local strain.
The remaining local strain aligns vorticity
with the intermediate eigenvector and does not
contribute significantly to amplification;
instead it surprisingly 
attenuates intense vorticity,
leading to breakdown of the observed power-law and
ultimately also the scale-invariance of vorticity amplification,
with important implications for prevailing intermittency theories.

\end{abstract}

\maketitle

%\paragraph{Introduction:}
Complex non-linear physical systems are often characterized
by formation of extreme events, which 
strongly deviate from Gaussianity, 
necessitating anomalous corrections to mean-field
descriptions \cite{Solli:2007,Rahmstorf:2011,sapsis_arfm}.
Fluid turbulence, described by 
the three-dimensional incompressible 
Navier-Stokes equations (INSE), is 
an emblematic example of such a system,
where extreme events are associated with 
intermittent formation of large velocity 
gradients, 
organized into thin filaments 
of intense vortices \cite{Siggia:81,Jimenez93,Ishihara09,BPBY2019}.
The amplification of such intense gradients 
is readily described by the vortex-stretching
mechanism, which expresses the non-linear stretching of
vorticity $\ww$, by the strain-rate tensor 
$S_{ij}$ in
the INSE (written as the vorticity equation):
\begin{align}
\frac{D \omega_i}{Dt} = \omega_j S_{ij} + \nu \nabla^2 \omega_i \ ,
\label{eq:vort}
\end{align} 
where $\nu$ is the kinematic viscosity.

The canonical description based on
angular momentum conservation dictates that
as vortical filaments are stretched by strain,
they become thinner and spin faster, enabling
gradient amplification, 
and simultaneously driving the energy cascade 
from large to small-scales \cite{tl72,carbone20}.
Though Eq.~\eqref{eq:vort} is valid pointwise,
this multiscale description 
can be analyzed by realizing 
that vorticity and strain are related
non-locally via Biot-Savart integral
over the entire flow domain:
\begin{align}
S_{ij} (\xx) = 
PV  \int_{\xx^\prime} 
\frac{3}{8\pi}  
(\epsilon_{ikl}r_j + \epsilon_{jkl}r_i)  \ \omega_l(\xx^\prime) \ 
\frac{r_k}{r^5} \ d^3 \xx^\prime \ ,
\label{eq:bs}
\end{align}
where $\rr=\xx - \xx^\prime$, $r=|\rr|$ and $\epsilon_{ijk}$
is the Levi-Civita symbol.
This integral essentially couples all the scales,
% in the flow,
providing a direct means to understand the non-locality of
gradient amplification,
without involving additional complexities
such as the pressure
field \cite{pumir94_pr,nomura:1998}.
%and supersedes spectral transfer analyses \cite{DR1990, mkv_book},
%which while successful at analyzing 
%global energy transfer characteristics, 
%cannot be applied to study extreme events.
However, the integral in Eq.~\eqref{eq:bs} is
analytically intractable, leading to outstanding challenges
in turbulence theory and also in establishing the regularity
of INSE \cite{doering2009}.
In this Letter, 
we investigate the nonlocality of 
vorticity self-amplification 
by tackling the Biot-Savart integral 
in Eq.~\eqref{eq:bs}
via direct numerical simulations (DNS) 
of INSE \cite{Ishihara09}.

To analyze the nonlocality w.r.t. a scale size $R$,
the integration domain in  Eq.~\eqref{eq:bs} 
is separated into a spherical neighborhood of radius $r \le R$, and
the remaining domain 
\cite{ham_pre08, ham_pof08,BPB2020}:
\begin{align}
S_{ij} (\xx) = 
\underbrace{ 
\int_{r>R}   \left[ \cdot\cdot\cdot \right]  d^3 \xx'}_{=S^{\rm NL}_{ij}(\xx,R)} 
\ + \
\underbrace{  
\int_{r \leq R}  \left[ \cdot\cdot\cdot \right] d^3 \xx'}_{=S^{\rm L}_{ij}(\xx,R)}  \ ,
\label{eq:lnl}
\end{align}
where $S^{\rm NL}_{ij}$  represents the non-local or background
strain acting on the vorticity to stretch it,
and $S^{\rm L}_{ij}$ is the local strain induced 
in response to stretching. 
We utilize DNS 
to compute $S_{ij}^{\rm L,NL}$ and investigate their interaction 
with vorticity for various $R$,
allowing us to quantify the degree of nonlocality
of vortex-stretching, and thereafter 
relate it to vortical structures in the flow.

\begin{figure*}
\begin{center}
\includegraphics[width=0.95\textwidth]{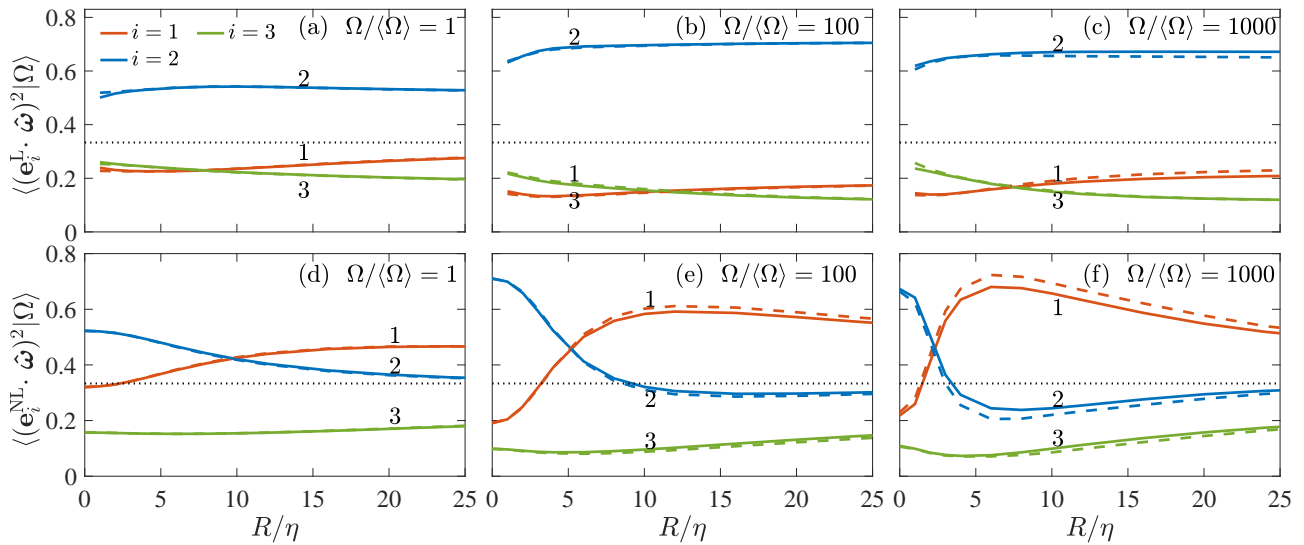}  \\
\vspace{-0.2cm}
\caption{
Conditional second moment of the alignment cosines
between vorticity and eigenvectors
of the local (L) and non-local (NL) strain tensors
at $\re=1300$ (solid lines) and $650$ (dashed lines),
and various conditioning values
of enstrophy $\Omega$.
The dotted line at $1/3$ in each panel corresponds
to a uniform distribution of the cosines.
Note, $S^{\rm L}_{ij}=0$ at $R=0$, with 
the alignments being undefined.
}
\label{fig:align}
\end{center}
\end{figure*}

While computing
$S_{ij}^{\rm L,NL}$ through numerical integration
is possible in DNS \cite{ham_pof08}, it is
prohibitively expensive 
at high Reynolds numbers. 
\footnote{for a $N^3$ grid, the
complexity of this integration scales as
$\mathcal{O}(N^6)$, making it practically unfeasible}.
Instead, as derived in our recent work \cite{BPB2020}, 
non-local (and local) strain
can be efficiently computed for any $R$
by applying a transfer function to the total strain
in Fourier space:
$\hat{S}_{ij}^{\rm NL} (\kk,R) = f(kR) \hat{S}_{ij} (\kk)$,
with
$f(kR) = {3 \left[ \sin(kR) - kR \cos(kR) \right] }/{(kR)^3}$,
thus bypassing
the direct evaluation of the Biot-Savart integral.
%\footnote{the complexity of this evaluation is
%same as that of 3D Fast-Fourier Transforms (FFTs),
%which is $\mathcal{O}(N^3 \log N)$}.
This novel approach is used to analyze
a large DNS database,
generated using the canonical setup of 
forced stationary isotropic turbulence 
in a periodic domain~\cite{Ishihara09},
utilizing the highly-accurate 
Fourier pseudo-spectral methods \cite{Rogallo}.
Special attention is given to maintain
a grid-resolution of smaller than 
the Kolmogorov length scale $\eta$ to 
resolve the extreme events accurately \cite{BPBY2019}.
The database 
corresponds to Taylor-scale Reynolds number $\re$
in the range $140-1300$, on up to grids
of $12288^3$ (for additional details see 
\cite{BS2020,BBP2020,BPB2020, BPB2021}).

%\paragraph{Alignment of vorticity and strain:}

The efficacy of vortex-stretching is 
controlled by the alignment between 
vorticity and strain-rate, and is 
commonly studied
%It is common to describe the alignments
in the eigenframe of strain tensor -- 
given by the eigenvalues $\lambda_i$ 
($\lambda_1 \ge \lambda_2 \ge \lambda_3$)
and the corresponding eigenvectors $\mathbf{e}_i$.
Incompressibility imposes 
$\lambda_1 + \lambda_2 + \lambda_3=0$,  
giving $\lambda_1>0$ and $\lambda_3<0$.
(The corresponding quantities for local/non-local strain 
are defined with superscripts L/NL).
It is well-known that $\lambda_2$ is
positive on average and vorticity
preferentially aligns with the
intermediate (second) 
eigenvector of the total strain rate 
\cite{Ashurst87,Tsi92,luethi:2005,BBP2020}.
This alignment is often regarded as anomalous, since
an analogy with stretching of material-lines suggests
that vorticity should align with the 
first eigenvector of total strain, corresponding
to the largest eigenvalue \cite{batchelor53}.

The earlier work of \cite{ham_pof08}, 
based on direct evaluation of the 
Biot-Savart integral for a single value of $R=12\eta$
at very low Reynolds number $\re\approx100$,
provides some evidence that vorticity preferentially
aligns with the first eigenvector of the 
non-local strain (similar to stretching
of material-lines), whereas
the anomalous alignment results from local dynamics. 
In the following, we 
provide a comprehensive investigation
of the alignment properties, as a function of
$R$ and over a drastically larger $\re$-range.
In addition, we also condition on
the enstrophy, $\Omega=\omega_i\omega_i$,
to analyze generation of intense vorticity.
To this end, we extract the second-moment
of directional cosines:
$\langle (\mathbf{e}_i^{\rm L,NL} \cdot \hat{\ww})^2\rangle$,
whose averages are individually bounded between $0$
and $1$ (with $1/3$ corresponding to a uniform
distribution), and additionally also
add up to unity, i.e.,
$\sum_{i=1}^3 (\mathbf{e}_i^{\rm L,NL} \cdot \hat{\ww})^2=1$
\cite{BBP2020}.

The directional cosines
%$\langle (\mathbf{e}_i^{\rm L,NL} \cdot \hat{\ww})^2\rangle$,
are shown as a function
of scale-size $R/\eta$ in
Fig.~\ref{fig:align}, and
conditioned on
$\Omega/\langle \Omega \rangle$ to
separate the extreme events. 
The alignments for $\SS^{\rm L}$
are explored first in Fig.~\ref{fig:align}a-c,
corresponding to 
$\Omega / \langle \Omega \rangle = 1, 100, 1000$.
We observe that for all $R/\eta$, vorticity
preferentially aligns with second eigenvector of 
$\SS^{\rm L}$, with a tendency to be
orthogonal to first and third eigenvectors.
The alignment properties become more pronounced 
as $\Omega$ increases.
Overall, this result conforms to the picture
of axisymmetric vortex tubes, where the velocity field
is approximately
two-dimensional, resulting
in preferential alignment of vorticity
with the second eigenvector of
$\SS^{\rm L}$ \cite{Jimenez92, moffatt94, ham_pre08}.
Interestingly,
vorticity is more orthogonal to the
first eigenvector compared to the third
for small $R$ ($\lesssim 10\eta)$, with the 
difference becoming
more pronounced for large $\Omega$ in panel c 
(we return to this behavior later).
At large $R$, this trend is reversed, 
approaching the well known result corresponding
to total strain (as $\SS^{\rm L}=\SS$
for $R\to\infty$) \cite{Ashurst87, BBP2020}.

The alignment of vorticity with $\SS^{\rm NL}$ 
is shown next in Fig.~\ref{fig:align}d-f.
The known alignment between vorticity and the intermediate
eigenvector of $\SS$ is recovered at $R = 0$
(where $\SS^{\rm NL}=\SS$).
However, as $R$ increases, 
a switch occurs and $\ww$ preferentially aligns
with the first eigenvector of $\SS^{\rm NL}$,
more strongly as $\Omega$ increases
(while vorticity is always preferentially orthogonal to 
third eigenvector)
\footnote{
Note, at $R\to\infty$, the non-local strain
vanishes, and hence all the alignments would just eventually 
approach $1/3$ corresponding to Gaussian background noise}.
These results clearly 
demonstrate that 
vortices are predominantly stretched by the non-local 
strain in a manner similar to passive material-lines,
with vorticity preferentially aligned with the 
most extensive eigenvector. 
However, in the vicinity of these vortices,
the (local) induced strain 
causes the alignment to switch
from first to second eigenvector.

%\textcolor{}{The analysis of Burgers vortices 
%%in~\cite{ham_pre08} 
%demonstrates that the value of $R$
%where the switching occurs, 
%see Fig.~\ref{fig:align}d-f, provides a good way to characterize the 
%filaments sizes~\cite{ham_pre08}.}

\begin{figure}
\begin{center}
\includegraphics[width=0.49\textwidth]{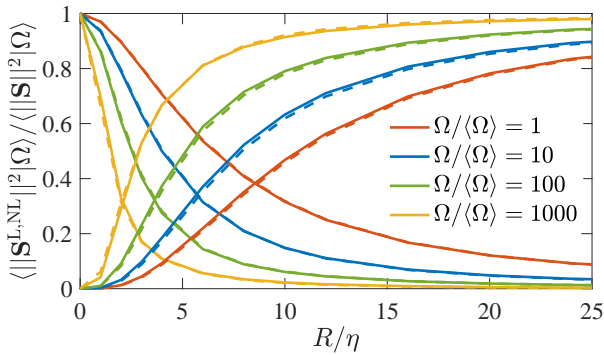} 
\caption{
Conditional expectation of the square-norm
of local (L) and non-local (NL) strain tensor, normalized by the
corresponding expectation of total strain, as a function
of $R/\eta$, at $\re=1300$ (solid lines)
and $\re=650$ (dashed lines). The curves for local
strain start from zero at $R=0$.
}
\label{fig:strain}
\end{center}
\end{figure}

\begin{figure}
\begin{center}
\includegraphics[width=0.45\textwidth]{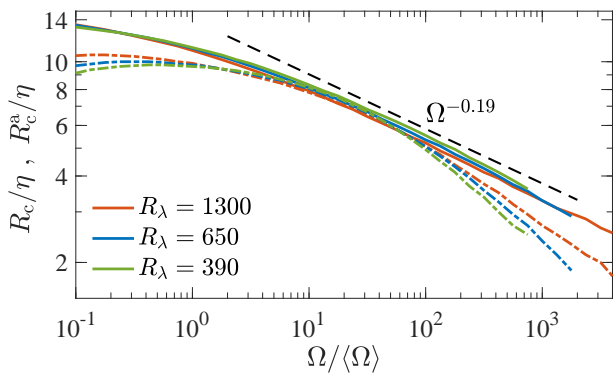}
\caption{
The critical distances $R_c/\eta$ (solid lines) and 
$R^a_c/\eta$ (dashed lines), as a function
of $\Omega$, respectively corresponding
to switching of alignment in Fig.~\ref{fig:align}d-f,
and the distance obtained from Fig.~\ref{fig:strain}
where magnitude of conditional local and non-local strain
are equal. 
The black dotted line corresponds $\Omega^{-0.19}$,
based on Eq.~\ref{eq:rom}, with $\gamma=0.76$ for $\re=1300$
\cite{BBP2020,supp}. 
}
\label{fig:rom1}
\end{center}
\end{figure}

%An important observation in 
Figure~\ref{fig:align}d-f shows that the switching of alignment
occurs at a distance $R^a_c = R^a_c (\Omega)$, which decreases
with $\Omega$.
This behavior also manifests itself when comparing the relative
magnitudes of $\SS^{\rm L, NL}$.
Fig.~\ref{fig:strain} shows the $R$-dependence of the conditional 
expectation of the norm of $\SS^{\rm L, NL}$.
They are normalized by the corresponding
conditional expectation of total strain,
which constrains the curves for $\SS^{\rm NL}$
and $\SS^{\rm L}$
at unity at $R=0$ and $\infty$ respectively.
As $\Omega$ increases,
the normalized magnitude of $\SS^{\rm L}$ approaches
unity at a smaller $R$, whereas that of 
$\SS^{\rm NL}$ falls of towards zero in a similar
fashion. This critical distance, say $R_c(\Omega)$, at which their 
relative magnitudes
are equal, steadily decreases with $\Omega$, 
qualitatively consistent with the switching
of alignment in Fig.~\ref{fig:align}d-f.

The results in Figs.~\ref{fig:align}-\ref{fig:strain}
allow us to identify characteristic length scales, which demarcate
the relative importance of local and non-local dynamics, 
and its dependence on $\Omega$.
The analysis of Burgers vortices
presented in \cite{ham_pre08},
establishes that that $R_c^a (\Omega) \simeq R_c (\Omega)$,
and they physically identify the radii of vortex tubes in
the flow \cite{ham_pof08}.
%From a structural point of view, it is natural
%to relate $R_c$ and $R^{a}_c$ to 
A simple method to obtain 
the radius of a vortex tube is
from a balance between viscosity $\nu$ and 
some effective strain $S$, giving 
$R = (\nu/S)^{1/2}$ \cite{Burgers48}.
Utilizing strain corresponding to mean-field,
i.e. $S \sim \langle \epsilon \rangle /\nu$,
where $\langle \epsilon \rangle$ is the mean-dissipation
rate, results in the well-known expression
for the Kolmogorov length scale 
$\eta= (\nu^3/ \langle \epsilon \rangle)^{1/4}$.
However, strain acting on intense vorticity
grows with vorticity, given by the 
power-law\cite{BPBY2019,BBP2020}:
\begin{align}
\langle ||\SS||^2 | \Omega \rangle \sim \Omega^\gamma \ , \ \ 0<\gamma<1
\label{eq:sigma}
\end{align}
where the exponent $\gamma$ weakly increases with 
$\re$, ostensibly approaching unity at $\re\to\infty$
\cite{supp}.
%\footnote{note this relation is only
%valid for events stronger than the mean value. For events weaker 
%than the mean value,
%strain and vorticity are seemingly uncorrelated,
%and the exponent $\gamma$ is zero;
%see \cite{supp}}. 
Utilizing Eq.~\eqref{eq:sigma},
and  $\langle \epsilon \rangle  = \nu \langle \Omega \rangle$
from statistical homogeneity,
the radius of tubes $R^*$ can be written 
as a function of $\Omega$:
\begin{align}
%{R_c(\Omega)/\eta \sim R_c^a(\Omega)/\eta} 
R^* /\eta \sim  (\Omega/\langle \Omega \rangle)^{-\gamma/4}  \ .
\label{eq:rom}
\end{align}
%Thus, it follows that as
%$\Omega$ increases, the vortex filaments becomes thinner.

To test the result in Eq.~\eqref{eq:rom}, 
Fig.~\ref{fig:rom1} shows the curves for
$R^a_c(\Omega)$ (dashed lines)
and $R_c (\Omega)$ (solid lines) 
extracted from 
from Fig.~\ref{fig:align}d-f 
and Fig.~\ref{fig:strain}, respectively.
Firstly, we observe that both 
$R_c(\Omega)$ and $R^a_c(\Omega)$
are always comparable and 
follow the same trend for moderately intense vorticity,
consistent with the power-law
predicted by Eq.~\eqref{eq:rom} (represented by the 
black dashed line). 
For very intense events ($\Omega/\langle \Omega \rangle \gtrsim 100$), 
$R_c(\Omega)$ is still consistent with the power-law, 
but $R^a_c(\Omega)$ starts deviating.
However, these deviations occur at slightly increasing 
values of $\Omega$ when $\re$ increases.
We note that over the range of $\re$
(from $390$ to $1300$),
the exponent $\gamma/4$ only varies
from $0.17$ to $0.19$ (respectively),
and this small change in slope is also faintly
visible for the curves corresponding to $R_c$. 
It is worth noting 
that such a dependence of vortex radius on $\Omega$
was not possible to 
detect in earlier studies at significantly 
lower $\re$ \cite{Jimenez93,ham_pof08}.

\begin{figure}
\begin{center}
\includegraphics[width=0.47\textwidth]{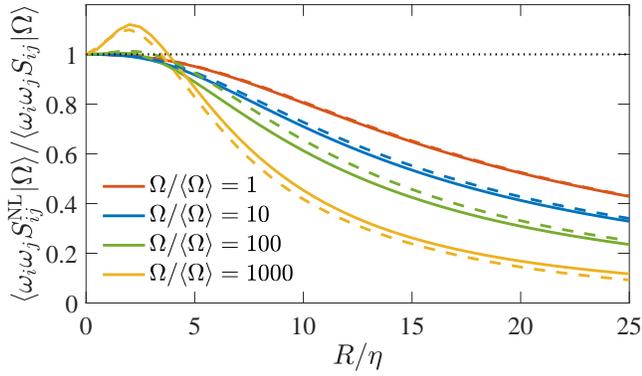} 
\caption{
Conditional expectation of the enstrophy production based
on non-local strain,
$\langle \omega_i \omega_j S^{\rm NL}_{ij} |\Omega \rangle$, 
normalized by the corresponding enstrophy production for total
strain, as a function of $R/\eta$, at $\re=1300$ (solid lines)
and $650$ (dashed lines). 
}
\label{fig:wsw}
\end{center}
\end{figure}

\begin{figure}
\begin{center}
\includegraphics[width=0.47\textwidth]{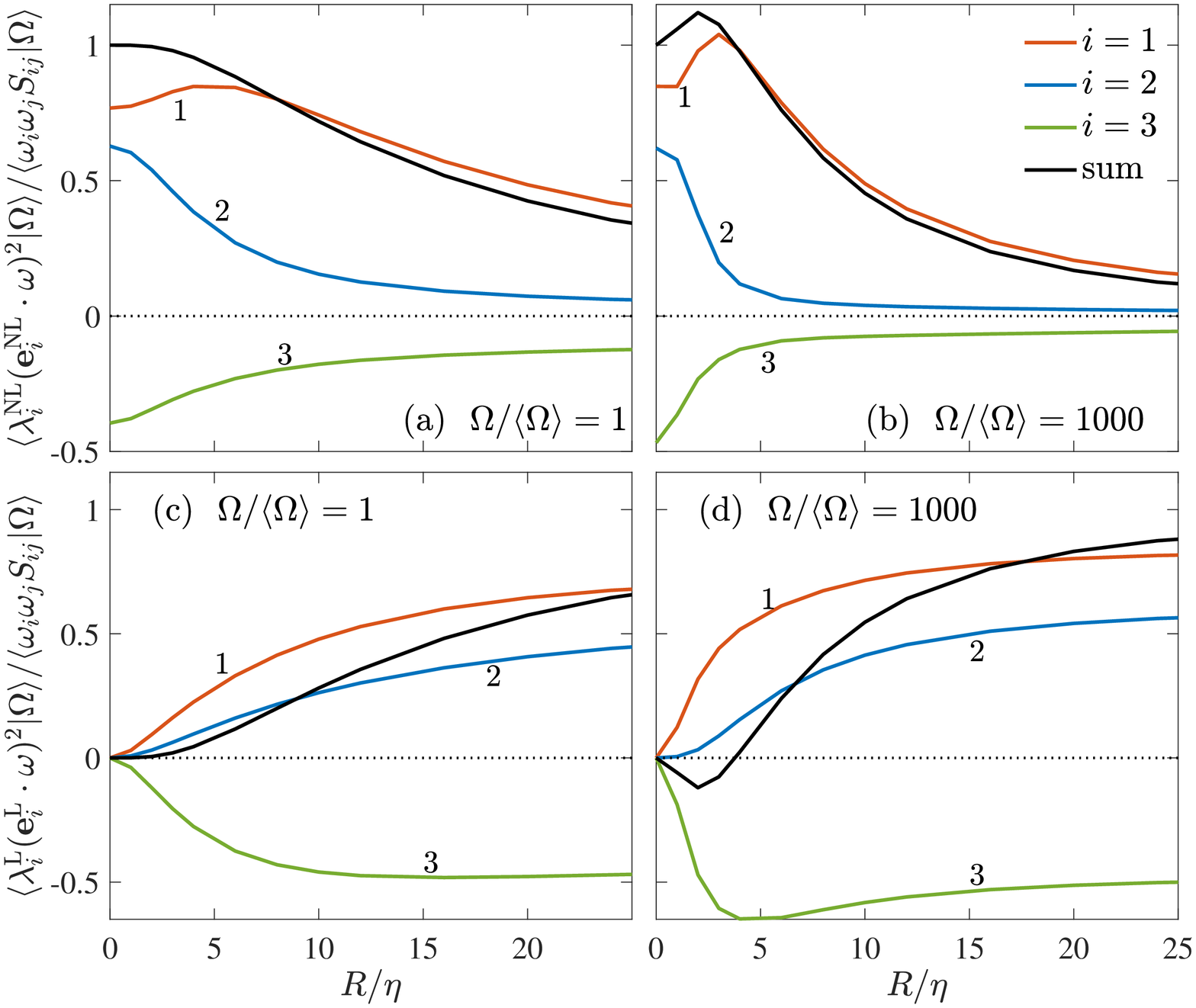}
\caption{
The individual contributions from each eigenvalue
to the non-local (NL) and local (L) enstrophy production terms,
normalized by the production based on total strain,
at $\re=1300$.
}
\label{fig:wew}
\end{center}
\end{figure}

To analyze deviations of $R^a_c$ at large $\Omega$,
%deviates from the expected power-law in Eq.~\eqref{eq:rom}
we consider the 
enstrophy production term,
$P_\Omega = \omega_i \omega_j S_{ij}$
%\footnote{The scalar term arising from 
%the vortex-stretching term in the
%enstrophy transport equation,
%as obtained by taking the dot product 
%of Eq.~\eqref{eq:vort} with $\omega_i$}, 
which also represents
the effective strain acting to amplify vorticity
by factoring in the alignments.
Similar to Eq.~\eqref{eq:lnl}, we can also
decompose  $P_\Omega$ as
$P_\Omega = P_\Omega^{\rm L} + P_\Omega^{\rm NL}$,
where $P_\Omega^{\rm L,NL} = \omega_i \omega_j S^{\rm L,NL}_{ij}$.
The conditional expectation of the 
non-local production $\langle P_\Omega^{\rm NL}|\Omega \rangle$,
normalized by the total conditional production 
$\langle P_\Omega|\Omega \rangle$,
is shown in Fig.~\ref{fig:wsw}.

For regions of moderately strong vorticity
($\Omega \lesssim 10 \langle \Omega \rangle$), 
the normalized production term $P_\Omega^{\rm NL}$
behaves qualitatively similar 
as non-local strain in 
Fig.~\ref{fig:strain} -- it starts at unity for $R=0$ 
and monotonically
decreases to zero at $R\to\infty$. However, when conditioned on
extreme values of $\Omega$ ($\gtrsim 100 \langle \Omega \rangle$),
the normalized $P_\Omega^{\rm NL}$ overshoots unity
at small $R$, before decreasing more sharply at larger $R$.
Since $P_\Omega^{\rm L}/P_\Omega = 1 - P_\Omega^{\rm NL}/P_\Omega$,
this observation 
implies that local production is negative for small $R$,
and thus counteracts vorticity amplification for large $\Omega$.
This is in fact 
a manifestation of the self-attenuation mechanism 
recently identified in \cite{BPB2020}, 
which provides an inviscid 
mechanism to arrest
vorticity growth and supports 
regularity of Navier-Stokes equations. 
Note, viscosity plays an implicit role,
since stationarity imposes a conditional balance
between net inviscid production and viscous
destruction, so that
the self-attenuation mechanism 
manifests at increasing $\Omega$ values 
with $\re$ \cite{BPB2020},
in agreement with the deviations
of $R^a_c(\Omega)$ in Fig.~\ref{fig:rom1}.

A breakdown of individual contributions from 
each eigenvalue for both $P_\Omega^{\rm L,NL}$,
normalized by the total production,
is shown next in Fig.~\ref{fig:wew}.
Fig.~\ref{fig:wew}a-b
shows that the first eigenvalue of non-local strain
provides most of the production, 
with the contributions
from the second and third eigenvalues
largely canceling each other;
except at small $R$, where the second eigenvalue 
provides a small but significant contribution.
The contributions to the local production
in Fig.~\ref{fig:wew}c-d
shows a very weak role of the intermediate eigenvalue
for small $R$, despite the very strong alignment observed in
Fig.~\ref{fig:align}a-c. Rather, 
the contributions from first and third eigenvalues
are more prominent, with the third eigenvalue ultimately
leading to overall negative local production at large $\Omega$
and small $R$ (which can also be traced to the slightly better
alignment of vorticity with the third eigenvector instead of the first,
also observed in Fig.~\ref{fig:align}a-c).
These results highlight the non-trivial role of nonlinearity,
going beyond a simple kinematic alignment switching 
as hypothesized earlier \cite{Jimenez92, ham_pre08}.
%$\Omega/\langle \Omega \rangle \gtrsim 100$. The weak change in the
%alignment property of $\ww$ with the eigenvalues of $\SS^{\rm {L, NL}}$ 
%leads to a major qualitative change in the vorticity amplification mechanism.

The results in Fig.~\ref{fig:wsw}-\ref{fig:wew} 
reiterate that vorticity
is predominantly amplified non-locally, analogous
to linear dynamics of material-line-stretching; whereas
the nonlinear effects are local 
and restricted to small distances,
but still playing an important role. 
Since as vorticity is amplified
beyond a threshold, the local effects
directly counteract
%directly start counteracting
further amplification, reflecting a fundamental
change in the nature of extreme events.
It marks
a breakdown of scale-invariance (self-similarity)
%change in the nature of extreme events.
%More precisely, it reveals a 
%breakdown of scale-invariance 
%(and hence
%self-similarity) 
of vorticity amplification at small-scales,
also explaining why the power-law
derived in Eq.~\eqref{eq:rom}
fails to capture the behavior of $R_c^a(\Omega)$
(in Fig.~\ref{fig:rom1})
for large $\Omega$. 
In contrast, for Burgers vortices,
for which $R_c^a(\Omega) = R_c(\Omega)$, the
stretching produced by local strain is always
zero \cite{ham_pre08}, i.e., the self-attenuation mechanism 
is always absent.

%Measuring the scale, $R_c^s(\Omega)$, defined by the condition
%that the production of vorticity from $\SS^{\rm L}$ and $\SS^{\rm NL}$ are
%comparable, provides another manifestation of the breaking down of 
%scale-invariance. 

\begin{figure}
\begin{center}
\includegraphics[width=0.47\textwidth]{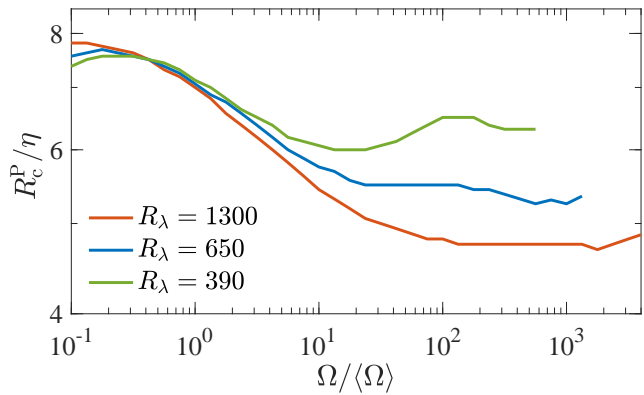}
\caption{
The critical distance
$R_c^P/\eta$ for which non-local 
enstrophy production accounts for
90\% of total production
(as derived from Fig.~\ref{fig:wsw}).
}
\label{fig:r_wsw}
\end{center}
\end{figure}

The breakdown of scale-invariance can further be shown 
by considering the critical scale $R_c^P = R_c^P (\Omega)$,
defined by the condition that non-local
enstrophy production recovers most of the
total production (as shown in Fig.~\ref{fig:r_wsw}).
%production term $P_\Omega$ recovers
%most of the total production term (as 
%derived from Fig.~\ref{fig:wsw}).
Remarkably, we find that
$R_c^P$ seemingly becomes constant at large $\Omega$,
marking a critical scale below
which the non-local effects do no penetrate
and local dynamics dominate.
A comparison with
Fig.~\ref{fig:rom1} shows that
the value of $R_c^P$ and range of $\Omega$ where its
constant are consistent
with where $R_c^a$ deviates from $R_c$ 
-- once again consistent with the onset of self-attenuation
mechanism \cite{BPB2020}.

The breakdown of scale-invariance (self-similarity) 
of vortex-stretching leads to some important
consequences for turbulence theory and modeling.
Prevalent intermittency theories postulate that
gradient amplification and the 
resulting energy-cascade is self-similar across scales, 
until regularized by viscosity. 
In fact, such an assumption is directly built into
celebrated Kolmogorov's hypotheses and also
multifractal and shell models \cite{Frisch95}.
However, current results point to an intricate
role of nonlinearity, which acts
in conjunction with viscosity to attenuate the most extreme events.
This casts serious doubts on the dimensional estimate of the scale 
where viscous effects 
become prevalent, as used by phenomenological models.
In fact, there is mounting evidence
that such models are inadequate at characterizing 
extreme events, even at large Reynolds numbers 
\cite{BPBY2019,BS2020, Elsinga:20}.
A similar situation also applies to large-eddy simulation,
where local dynamics are unresolved (by definition).
The current results call for development of new models
which can, for instance, appropriately capture the self-attenuation
mechanism.

In conclusion, using state-of-the-art DNS,
we have analyzed non-locality
of vorticity-amplification by directly tackling the
global Biot-Savart integral. 
%which couples vorticity
%and strain-rate.
We show that
vorticity is predominantly
amplified by non-local strain, with the underlying dynamics
being linear. 
We identify the characteristic scale of nonlocality, 
which varies as a simple power-law of vorticity magnitude.
The nonlinear effects are
captured by the remaining local strain, revealing
that the nature of extreme events is fundamentally different 
due to the self-attenuation mechanism~\cite{BPB2020}, 
ultimately leading to a 
breakdown of the observed power-law and 
scale-invariance of vortex-stretching mechanism.
Further investigations
are ongoing and are expected to 
provide essential ingredients for
improved intermittency theories and 
turbulence models.

\begin{acknowledgments}

\paragraph{Acknowledgements:}
We gratefully acknowledge the Gauss Centre for Supercomputing e.V. for 
providing computing time on the supercomputers JUQUEEN and
JUWELS at J\"ulich Supercomputing Centre (JSC), 
where the simulations reported in this
paper were performed.  

\end{acknowledgments}

%\bibliography{large_grad}


\end{document}